\documentclass[aip,rsi,reprint]{revtex4-1}

\usepackage{amsmath,amssymb,bm,graphicx}
\usepackage{xcolor}
\usepackage[colorlinks=true,urlcolor=blue,linkcolor=blue,citecolor=blue,bookmarks=false]{hyperref}

\begin{document}

\title{Ultra-high vacuum compatible preparation chain for intermetallic compounds}

\author{A. Bauer}
\email{andreas.bauer@frm2.tum.de}
\affiliation{Physik-Department, Technische Universit\"{a}t M\"{u}nchen, D-85748 Garching, Germany}

\author{G. Benka}
\affiliation{Physik-Department, Technische Universit\"{a}t M\"{u}nchen, D-85748 Garching, Germany}

\author{A. Regnat}
\affiliation{Physik-Department, Technische Universit\"{a}t M\"{u}nchen, D-85748 Garching, Germany}

\author{C. Franz}
\altaffiliation{present address: Heinz Maier-Leibnitz Zentrum (MLZ), D-85748 Garching, Germany}
\affiliation{Physik-Department, Technische Universit\"{a}t M\"{u}nchen, D-85748 Garching, Germany}

\author{C. Pfleiderer}
\affiliation{Physik-Department, Technische Universit\"{a}t M\"{u}nchen, D-85748 Garching, Germany}

\date{\today}

\begin{abstract}
We report the development of a versatile material preparation chain for intermetallic compounds that focuses on the realization of a high-purity growth environment. The preparation chain comprises of an argon glovebox, an inductively heated horizontal cold boat furnace, an arc melting furnace, an inductively heated rod casting furnace, an optically heated floating-zone furnace, a resistively heated annealing furnace, and an inductively heated annealing furnace. The cold boat furnace and the arc melting furnace may be loaded from the glovebox by means of a load-lock permitting to synthesize compounds starting with air-sensitive elements while handling the constituents exclusively in an inert gas atmosphere. All furnaces are all-metal sealed, bakeable, and may be pumped to ultra-high vacuum. We find that the latter represents an important prerequisite for handling compounds with high vapor pressure under high-purity argon atmosphere. We illustrate operational aspects of the preparation chain in terms of the single-crystal growth of the heavy-fermion compound CeNi$_{2}$Ge$_{2}$.
\end{abstract}

\vskip2pc

\maketitle

\section{Introduction}

High-quality single crystals are an important prerequisite for major advance in experimental solid state research. Here, high quality refers to a high degree of structural order with a small crystalline mosaicity, low concentrations of defects and impurities, and the absence of parasitic phases. In order to prepare samples that satisfy these criteria, it is crucial to avoid contaminations in every step of the preparation process. The latter often starts with the synthesis of polycrystalline material. The resulting specimens are either directly investigated regarding their physical properties or processed further into single crystals. In either case, highest possible sample purity is essential to ensure that the intrinsic behavior of the material may be studied. For instance, in materials with strong electronic correlations high sample purity has been a precondition for the discovery of a variety of unexpected properties including unconventional forms of superconductivity,\cite{2007:Monthoux:Nature, 2009:Pfleiderer:RevModPhys, 2011:Stewart:RevModPhys} new types of spin order and spin excitations,\cite{2007:Borzi:Science, 2008:Castelnovo:Nature, 2009:Muhlbauer:Science, 2010:Balents:Nature} as well as non-Fermi liquid behavior near magnetic quantum phase transitions.\cite{2001:Stewart:RevModPhys, 2007:Lohneysen:RevModPhys} Many of these phenomena are very sensitive to defects such as vacancies, interstitial and substitutional impurities, antisite disorder, dislocations, or grain boundaries.

In intermetallic compounds the high affinity of the starting elements to oxygen, nitrogen, and hydrocarbons poses a great challenge for sample quality. This problem is especially pronounced for many rare-earth elements that quickly oxidize when exposed to ambient conditions. In addition, complexities due to high vapor pressures, peritectic formations, or wide solubility ranges are observed, similar to other materials classes. Common techniques for the preparation of polycrystalline samples of intermetallic compounds include solid state reactions, arc melting, resistive melting, and inductive melting. The resulting polycrystals are often treated by an annealing process or used for single-crystal growth by means of float-zoning with inductive or optical heating. Single crystals may also be synthesized directly from the starting elements using Czochralski growth with tetra-arc or induction heating, (flux) Bridgman growth, or chemical vapor transport.

Optical float-zoning, as our method of choice, offers several advantages: (i)~it is crucible-free, (ii)~impurities usually possess a solubility in the molten state that is considerably larger than in the solid state hence accumulating in the zone instead of being incorporated into the growing crystal, (iii)~it may be carried out on insulating as well as conducting materials, (iv)~it allows to address compounds with comparatively high vapor pressures as only a small amount of material is molten at a given time, (v)~it allows to monitor the molten zone during the growth permitting to systematically optimize the growth parameters, and (vi)~it allows to prepare, in principle, peritecticly forming compounds by means of traveling solvent growth. In order to benefit from these advantages float-zoning requires reasonably shaped feed rods with the appropriate stoichiometry, high compositional homogeneity, and sufficient mass density.

To ensure the highest sample quality possible, we have developed a single-crystal preparation chain that is optimized for intermetallic compounds. Starting from high-purity elements, this chain allows us to prepare polycrystalline ingots for a wide range of materials while handling air-sensitive elements exclusively under inert atmosphere. Subsequently, polycrystalline rods may be cast and float-zoned followed by post-growth heat treatment. The chain comprises of (i)~an argon glovebox with a custom-made load-lock, (ii)~an arc melting furnace, (iii)~an inductively heated horizontal cold boat furnace, (iv)~an inductively heated rod casting furnace, (v)~an optically heated floating-zone furnace, (vi)~a resistively heated annealing furnace, and (vii)~an inductively heated annealing furnace. The furnaces are all-metal sealed, bakeable, and may be pumped to ultra-high vacuum. For most materials, due to high vapor pressures of the elements involved, the ultra-high vacuum represents a precondition for the application of a high-purity inert atmosphere consisting of 6N argon additionally passed through point-of-use gas purifiers.\cite{SAESgetter}

In this paper, we report the operation of the entire preparation chain and explain specific aspects of the setups not reported before. We describe the development of the arc melting furnace and the cold boat furnace, both of which may be docked to the glovebox allowing to handle air-sensitive elements in an inert environment. For the cold boat furnace, the docking process required the design of a bespoke metal bellows load-lock. We report further the design of two annealing furnaces based on a commercial Knudsen-type effusion cell and inductive heating, respectively, permitting the heat treatment of both polycrystalline and single-crystalline material. Detailed descriptions of the rod casting furnace and our refurbished image furnace have been reported in Refs.~\onlinecite{2016:Bauer:RevSciInstrum} and \onlinecite{2011:Neubauer:RevSciInstrum}, respectively.

Our paper is organized as follows. In Sec.~\ref{development} we describe the main technical developments in the following order: argon glovebox, arc melting furnace, cold boat furnace, metal bellows load-lock, rod casting furnace, optical floating-zone furnace, resistively and inductively heated rod casting furnace. In Sec.~\ref{growth} we illustrate the typical use of the preparation chain for the example of the single-crystal growth of CeNi$_{2}$Ge$_{2}$. This rare-earth compound has been reported to be close to an antiferromagnetic quantum critical point at ambient pressure and its physical properties depend sensitively on sample quality making it an ideal benchmark for crystal growth. Finally, we summarize the main developments and results in Sec.~\ref{conclusions}.

\section{Technical developments}
\label{development}

\subsection{Argon glovebox}

\begin{figure}
\includegraphics[width=1.0\linewidth]{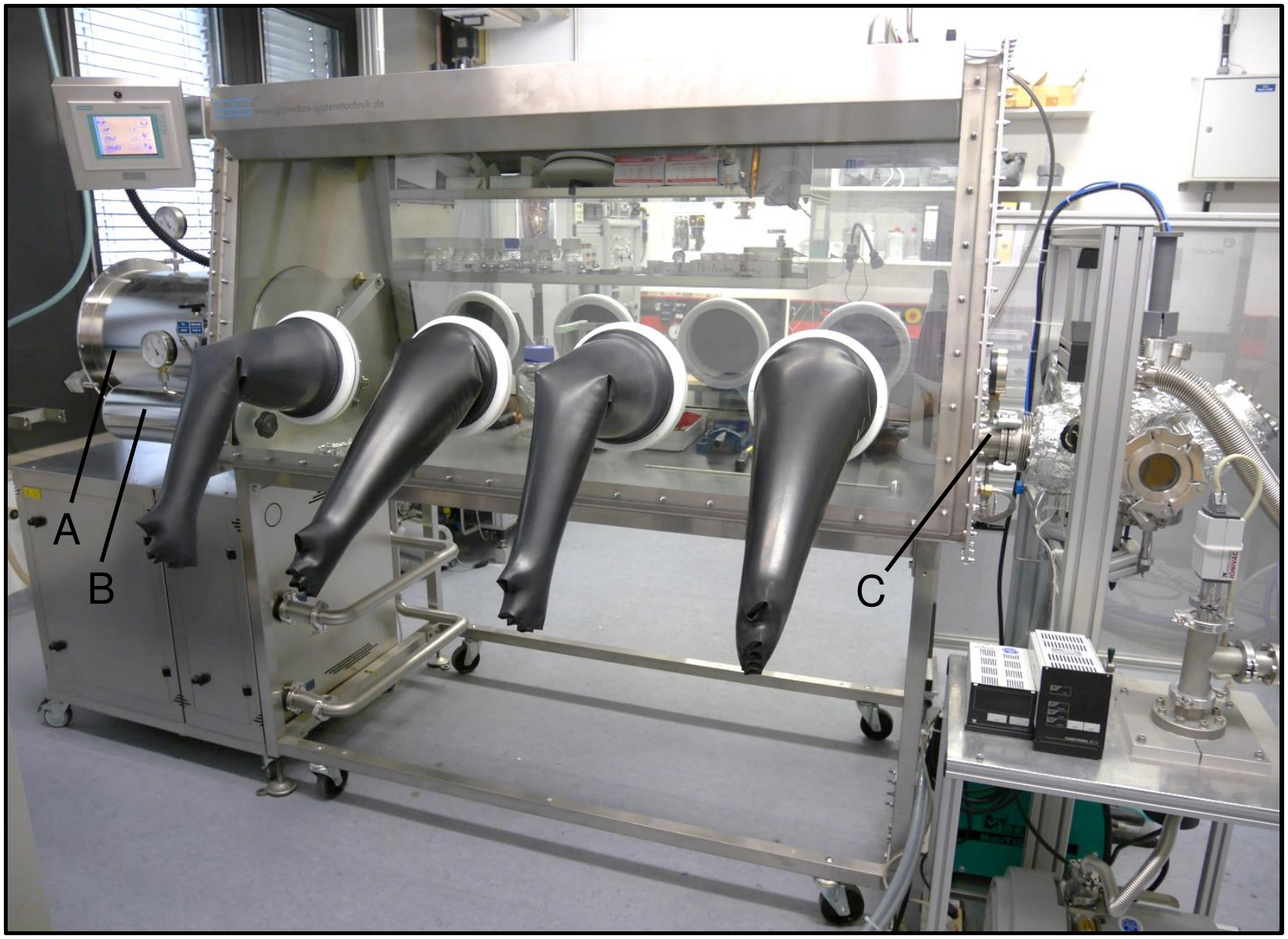}
\caption{(Commercial argon glovebox from GS as equipped with various bespoke components. The glovebox includes three locks (A), (B), and (C). On the right-hand side, the arc melting furnace is connected to the glovebox via lock~(C).}
\label{figure01}
\end{figure}

The starting point of our material preparation chain is a commercial glovebox from GS\cite{GS}, shown in Fig.~\ref{figure01}, that has been designed to suit our purposes. This glovebox provides an argon atmosphere containing about 1~ppm oxygen and 1~ppm water. Eight gloves, four on either side, provide the necessary access to the working space. An automatic lock with a diameter of 400~mm~(A) and a manual lock with a diameter of 150~mm~(B) are located on one side (left on photograph). A third lock is mounted on the opposite side~(C). This lock consists of an ISO-K100 flange of 70~mm length and a lid that is tightened manually. A manometer and a three-way valve permit to purge the system. The arc melting furnace or the horizontal cold boat furnace may be attached to the glovebox by means of this third lock. For the arc melting furnace a CF DN100 plate valve is used in combination with a short U-profile bellows for length adaptation. The cold boat is connected via a bespoke metal bellows stabilized by telescope rods (the docking process is described in detail below).

\subsection{Arc melting furnace}

\begin{figure}
\includegraphics[width=1.0\linewidth]{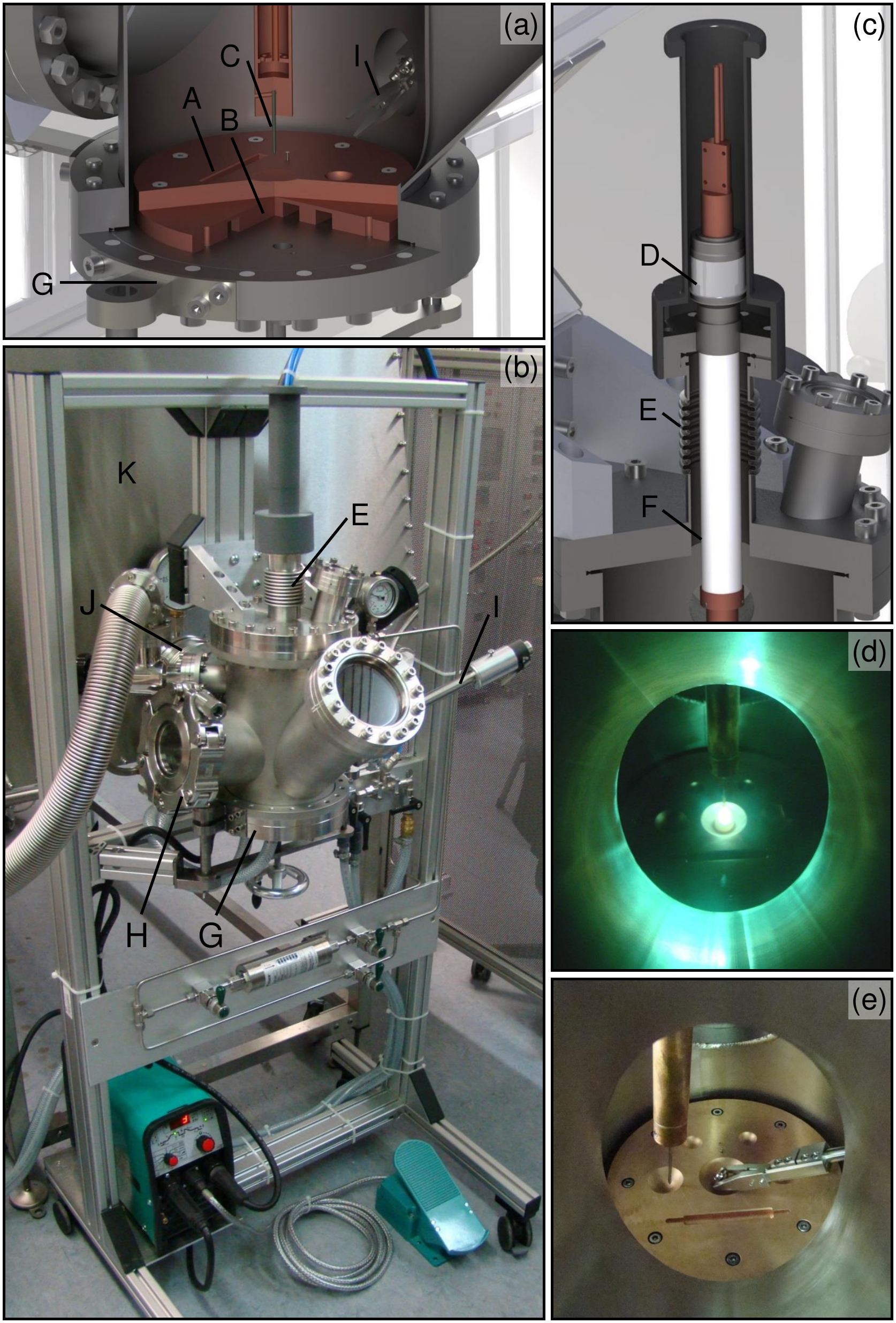}
\caption{All-metal sealed arc melting furnace. (a)~Cut-away view of the lower part of the vacuum chamber. (b)~Total view of the arc melting furnace docked to the glovebox. (c)~Cut-away view of the top flange. (d)~View through the front window during melting. (e)~Flipping a sample using the wobble stick. See text for details.}
\label{figure02}
\end{figure}

Arc melting furnaces are widely used for the preparation of polycrystalline samples of intermetallic compounds. When arc melting a sample, heating is carried out by an argon plasma that is ignited between a tungsten tip electrode and a base plate at earth potential. The process compares somewhat to tungsten inert gas welding. Standard viton-sealed arc melting furnaces are commercially available in various sizes and designs. For our studies, we have developed an all-metal sealed setup that may be loaded from the glovebox, as depicted in Fig.~\ref{figure02}.

Central part of the arc melting furnace is a copper plate~(A) providing molds of different sizes and shapes. This plate is tightly bolted to a water-cooled CF DN160 flange~(B) at the bottom of a bespoke vacuum chamber. The electrode consists of a lanthanum doped tungsten tip~(C) attached to a water-cooled current feedthrough~(D). Horizontal and vertical movement of the electrode is made possible by a DN40 U-profile bellows welded to the CF DN160 flange at the top of the chamber~(E). The electrode is electrically insulated by Stycast~2762 and a sleeve made of polyether ether ketone~(PEEK)~(F). Electrical power is supplied by a Merkle MobiTIG 190 DC welding transformer.\cite{Merkle} The bottom flange of the system is mounted on a support that may be lowered using a crank handle and pivoted to the front~(G). This way, easy access is provided for cleaning of the vacuum chamber, loading and unloading of samples, as well as replacing the copper plate or the tip of the electrode.

A VACOM Quick CF DN100 port\cite{Vacom} tightened by a clamping chain located on the left side of the vacuum chamber~(H) permits rapid access, e.g., for sample loading. Windows at the front (CF DN100) and the top (CF DN40) of the vacuum chamber as well as in the Quick CF flange allow to monitor the melting process. A pyrometer may be attached to the window at the top. Sample manipulation is made possible by means of a wobble stick on the right-hand side of the vacuum chamber~(I). Thus, without opening the system, one may flip alloyed pills and collect fragments that may have been created during the heating or cooling of the sample.

A CF DN100 plate valve at the back of the vacuum chamber~(J) allows to connect the arc melting furnace to the glovebox~(K), where a short U-profile bellows with one CF DN100 and one ISO-K DN100 flange serves for adaptation of the length. Purging is provided via the glovebox. Subsequently, the plate valve is opened and the elements are loaded from the glovebox with a bespoke shovel. The wobble stick may be used to move sample pieces into the molds of the copper plate.

After closing the plate valve or, alternatively, after loading a sample from ambient, the system is pumped by an external turbomolecular pump backed by an oil-free scroll pump. When baking the arc furnace using permanently installed heating tapes, pressures in the range of $10^{-8}$~mbar are reached within a day. Finally, the chamber is flooded with up to 1.2~bar of high-purity argon. Other than the cold boat system, the arc melting furnace may stay docked to the glovebox during the melting process which permits a high sample output even when handling elements sensitive to air or moisture. As an empirical observation, we note that the plasma arc is very stable in a high-purity argon atmosphere.\cite{2003:Boeuf:PhD} In turn, the furnace may be operated at much reduced electrical power which allows to heat the samples in a more gentle and controlled manner as compared to regular arc melting furnaces.

Typically, the arc melting furnace is used for the first synthesis of compounds of interest. If the constituents react forming an ingot with homogenous appearance, a subsequent analysis by means of polished cuts and X-ray powder diffraction provides information on the metallurgical phase diagram and the chances of success of single-crystal growth. The arc melting furnace is also very useful in studies that require a large number of high-quality polycrystalline samples, e.g., extended doping series, as well as studies involving high-melting or insulating compounds. Bespoke copper plates with elongated or large-diameter cylindrical molds permit the direct preparation of feed rods for float-zoning or targets for thin film growth by means of sputtering or pulsed laser deposition, respectively. Last but not least, the furnace may also be used to seal metallic ampules under high-purity inert atmosphere.

\subsection{Horizontal cold boat furnace}

\begin{figure}
\includegraphics[width=1.0\linewidth]{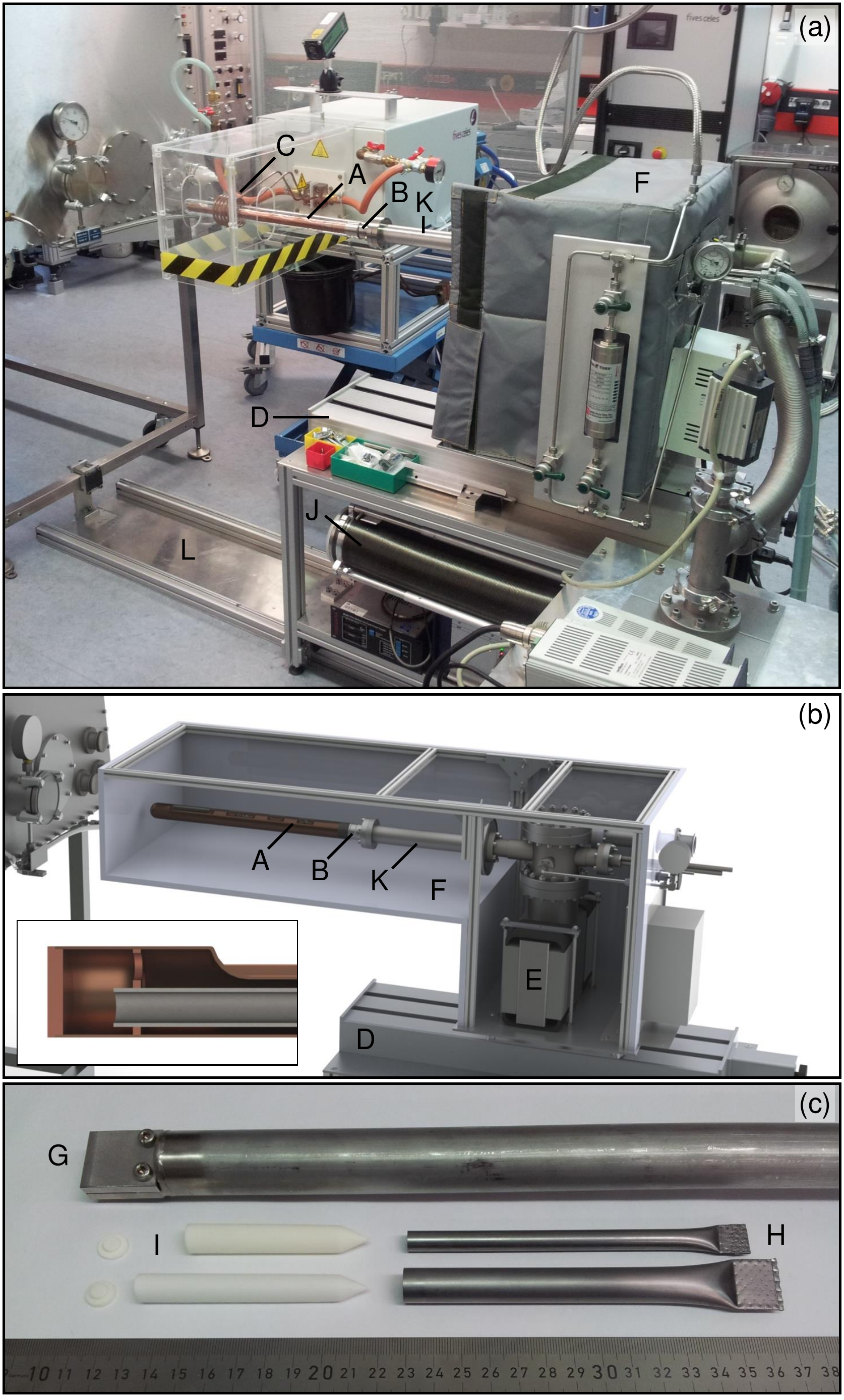}
\caption{All-metal sealed horizontal cold boat furnace. (a)~Total view including the permanently installed part of the heating tent and the gas purification system. The custom lock of the glovebox is visible on the left. A two-part guide rail on the floor provides the room to maneuver for the docking process. See text for details. (b)~Isometric view. The inset shows a cut-away view of the tip of the water-cooled cold boat. (c)~Water-cooled pulling rod for horizontal Bridgman growth with tantalum ampules and ceramic crucibles.}
\label{figure03}
\end{figure}

The horizontal cold boat system described in this section represents a versatile and simple setup which, in combination with a radio frequency~(RF) generator, is typically used to heat and melt metallic specimens. The system developed for our work is depicted in Figs.~\ref{figure03}(a) and \ref{figure03}(b). Central part is a water-cooled copper hearth~(A) with four molds of different sizes. Samples are placed in these molds either directly or in crucibles made of refractory metals, ceramics, or quartz glass. After closing and pumping the system, the samples are heated by means of RF induction. Water cooling prevents the cold boat from heating up thereby reducing potential contamination during sample preparation.

In order to achieve sufficient heating for the synthesis of typical intermetallic compounds at least one of the constituents should be metallic and of a size in the millimeter range. Metallic susceptors, however, permit to handle also semiconductors such as silicon. The radiant heat of the susceptor heats the specimen increasing the charge carrier concentration to a value at which the RF induction couples sufficiently to the material.\cite{1994:Tautz:PhD} Moreover, crucibles made of tungsten, tantalum, or molybdenum allow for the indirect heating of powder or insulating samples, as for instance demonstrated by the preparation of transition metal diborides from high-purity powder educts by means of a solid state reaction.\cite{2014:Bauer:PhysRevB} 

Besides the large range of materials that may be treated, the well-controlled heating and the simple design represent important advantages of the cold boat system. For our work, we realized a compact and all-metal sealed setup on top of an ion getter pump without moveable parts in the vacuum chamber. Samples are loaded and unloaded by removing a quartz-glass tube attached to a standard CF DN40 quartz-to-metal seal~(B). The coil~(C) of the RF oscillator circuit of a Fives Celes MP 50~kW RF generator\cite{FivesCeles} surrounds the quartz-glass tube. Horizontal movements of the crucible with respect to the RF coil is made possible by a linear actuator~(D) supporting the entire vacuum chamber.

An external turbomolecular pump backed by an oil-free scroll pump is used prior to the operation of the Duniway V60 diode ion getter pump~(E).\cite{Duniway}. In combination with a custom-built two-part heating tent~(F), of which only the part covering the cold boat has to be removed during the melting process, we reach final pressures of the order of $10^{-10}$~mbar. When handling compounds with high vapor pressure, the chamber is operated under an high-purity argon atmosphere at a pressure of up to 1.5~bar. Getter materials such as titanium sponges or rare-earth metal pills may be placed in one of the smaller molds and heated prior to melting the sample to purify the argon gas.

In order to further enhance the versatility of the setup, the cold boat described above may be replaced by a water-cooled horizontal Bridgman system~(G), see Fig.~\ref{figure03}(c). Here, tantalum ampules~(H) are used as hot crucibles either as such or in combination with crucibles~(I) made of alumina (Al$_{2}$O$_{3}$) or zirconia (ZrO$_{2}$). Temperatures are monitored by an IMPAC IGA 140 MB 25 L pyrometer.\cite{LumaSense}

\subsection{Metal bellows load-lock}

\begin{figure}
\includegraphics[width=1.0\linewidth]{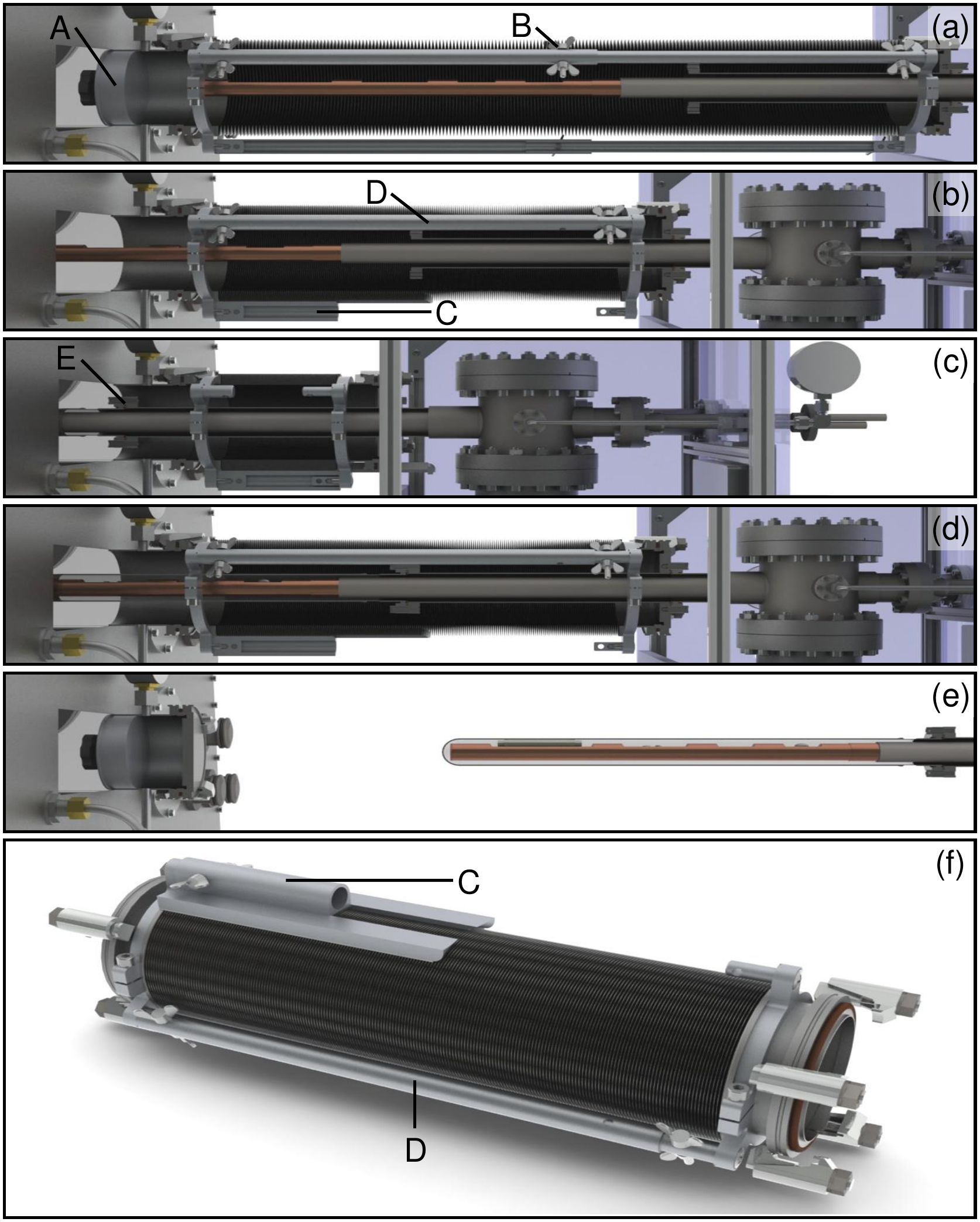}
\caption{Side view of the docking process between the glovebox and the cold boat furnace. \mbox{(a)--(e)}~Five steps within the docking process. See text for details. (f)~Isometric view onto the bottom side of the metal bellows lock corresponding to steps (b) and (d), respectively.}
\label{figure04}
\end{figure}

While many compounds may be loaded into the cold boat system at ambient conditions, special care is required when elements are processed that are highly sensitive to oxygen or moisture. For instance, high-purity forms of rare-earth elements such as cerium or europium are commercially available and shipped in glass or stainless steel ampules sealed under argon atmosphere. In order to avoid contamination, these elements should be handled exclusively in an inert atmosphere either until they are alloyed into a substance that is stable in air or until they are prepared for measurements. For binary compounds, typically the target material is directly synthesized. For ternary and quaternary samples, we frequently prepare a binary alloy containing the amount of the reactive element necessary for the target material and a second constituent that is easy to handle.

To satisfy these criteria, we have developed a metal-bellows-based load-lock system that connects the cold boat furnace, cf.\ Fig.~\ref{figure03}, to the glovebox in order to load the samples under inert conditions. Subsequently, the cold boat system is closed under the argon atmosphere of the glovebox before it is detached from the latter and pumped to ultra-high vacuum. Following this procedure,  samples are treated the same way independent whether they were loaded from the glovebox or at ambient.

The load-lock~(J) consists of a highly flexible metal bellows with ISO-K100 flanges that is stabilized by telescope rods and a sled, respectively. The metal bellows is only mounted for the actual docking process and otherwise stored in the rack of the cold boat furnace, as depicted in Fig.~\ref{figure03}(a). An adapter flange~(K), which connects the load-lock with the cold boat system and provides the necessary headroom for the compressed bellows, is permanently installed on the cold boat system. The adapter is located between the quartz-to-metal seal and the central vacuum chamber. The horizontal translation necessary for the docking process exceeds the capacity of the linear actuator that is used to position the cold boat during the melting processes. Therefore, the entire cold boat system including the linear actuator is placed on top of a guide rail~(L) and may be moved manually. 

In the following we describe the docking procedure in detail. Fig.~\ref{figure04}(a) shows a cut-away view with the glovebox on the left-hand side and the cold boat system without the quartz-glass tube on the right-hand side. They are connected through the bespoke metal bellows fixed at maximum elongation. The lock of the glovebox is sealed by a cap~(A). The metal bellows as well as the cold boat system are at ambient. Using the pumping systems of the glovebox, the metal bellows and the cold boat system are evacuated and subsequently flooded with argon at least three times. After this purging, the cap and the fixations of the metal bellows~(B) are removed as depicted in Fig.~\ref{figure04}(b). Next, the cold boat system is moved towards the glovebox until the metal bellows is compressed halfway. In this position, the lower rod of the telescope rack of the metal bellows is replaced by a sled~(C). As the next step, illustrated in Fig.~\ref{figure04}(c), the two upper rods of the telescope rack~(D) are removed and the metal bellows is fully compressed.

Now, the cold boat is loaded accessing it from the inside of the glovebox and the quartz-to-metal seal~(E) is tightened while the cold boat system is filled with the argon of the glovebox. Subsequently, cf.\ Fig.~\ref{figure04}(d), steps (b) and (a) are repeated in reverse order. Following the removal of the metal bellows load-lock, depicted in Fig.~\ref{figure04}(e), the cold boat system is ready to be pumped to ultra-high vacuum.

In our crystal growth chain the cold boat system with its bespoke metal bellows load-lock is mostly used for the initial synthesis of compounds that are stable at ambient while involving air-sensitive elements. We note that, despite the stirring of the molten sample caused by the RF induction, the resulting specimens may exhibit insufficient compositional homogeneity arising from the temperature gradient between the top of the sample and the bottom close to the water-cooled crucible. This disadvantage, however, may either be overcome by turning the specimen upside down and remelting it several times or by processing it further in the rod casting furnace described below.

\subsection{Rod casting furnace}

\begin{figure}
\includegraphics[width=1.0\linewidth]{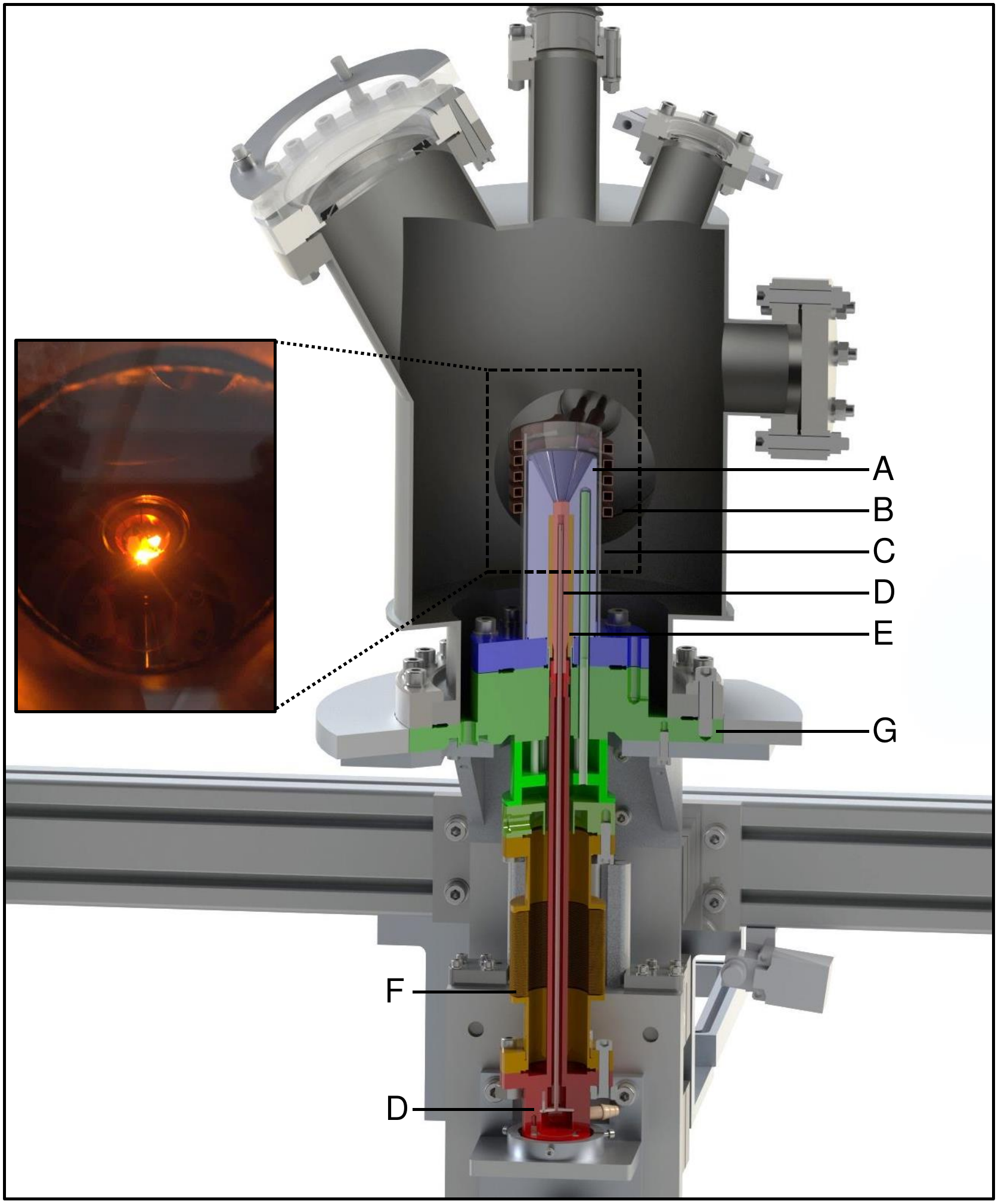}
\caption{All-metal sealed rod casting furnace in cut-away view. (A)~Hukin type crucible, (B)~induction coil, (C)~quartz glass tube, (D)~pulling rod, (E)~casting mold, (F)~highly flexible metal bellows, (G)~support flange providing water cooling. The inset shows the view through one of the windows during sample heating.}
\label{figure05}
\end{figure}

One of the main prerequisites of single-crystal growth by means of float zoning is the preparation of well-shaped and mechanically stable feed rods with well-defined stoichiometry, high compositional homogeneity, and sufficient mass density. In particular when preparing rods in the form of compressed or sintered powders, high packing density represents an important technical challenge. When casting the rods from the melt, they should not contain cavities. To meet these requirements, we designed an all-metal sealed inductively heated rod casting furnace, as reported in detail in Ref.~\onlinecite{2016:Bauer:RevSciInstrum}. In the following, we briefly summarize key aspects of its design and operation for completeness.

As shown in Fig.~\ref{figure05}, central part of the furnace is a water-cooled Hukin type crucible made of copper~(A) in which the educts are placed. The sample is heated by means of RF induction using a built-in coil~(B). A quartz glass tube serves as electrical insulation~(C). A water-cooled pulling rod~(D) completes the sample space. In order to ensure excellent compositional homogeneity, the sample is typically molten and cooled down several times where the solid ingot may be flipped by the sudden upwards movement of the pulling rod. Then, the sample is molten a final time and the pulling rod is moved downwards permitting the melt to flow into the casting mold~(E) where a polycrystalline rod forms. A metal bellows~(F) provides the necessary vertical movability.

The crucible is operated in an all-metal sealed vacuum chamber. For sample changes, the CF DN160 flange of the central support flange~(G) is opened and the lower section of the casting furnace is moved downwards using a linear actuator. Subsequently, the crucible is untightened and removed in order to access the mold containing the polycrystalline rod. Note that the tip of the pulling rod is exchangeable which permits, in combination with bespoke casting molds, to modify the furnace within minutes for the preparation of rods with diameters between 6~mm and 10~mm and a length up to 90~mm.

Prior to sample preparation, the system is evacuated using a turbomolecular pump backed by an oil-free scroll pump. In combination with permanently installed heating tapes and a heating jacket for the metal bellows, we reach final pressures of $10^{-8}$~mbar before applying a high-purity argon atmosphere of up to 3~bar.

Besides the casting of polycrystalline feed rods for float-zoning, we also use this furnace for the purification of elements. When starting elements exhibit oxygen contaminations, such as commercially available manganese, we find that solid oxide flakes tend to float on top of the molten metallic sample and remain as slag in the crucible after casting a rod.

\subsection{Optical floating-zone furnace}

\begin{figure}
\includegraphics[width=1.0\linewidth]{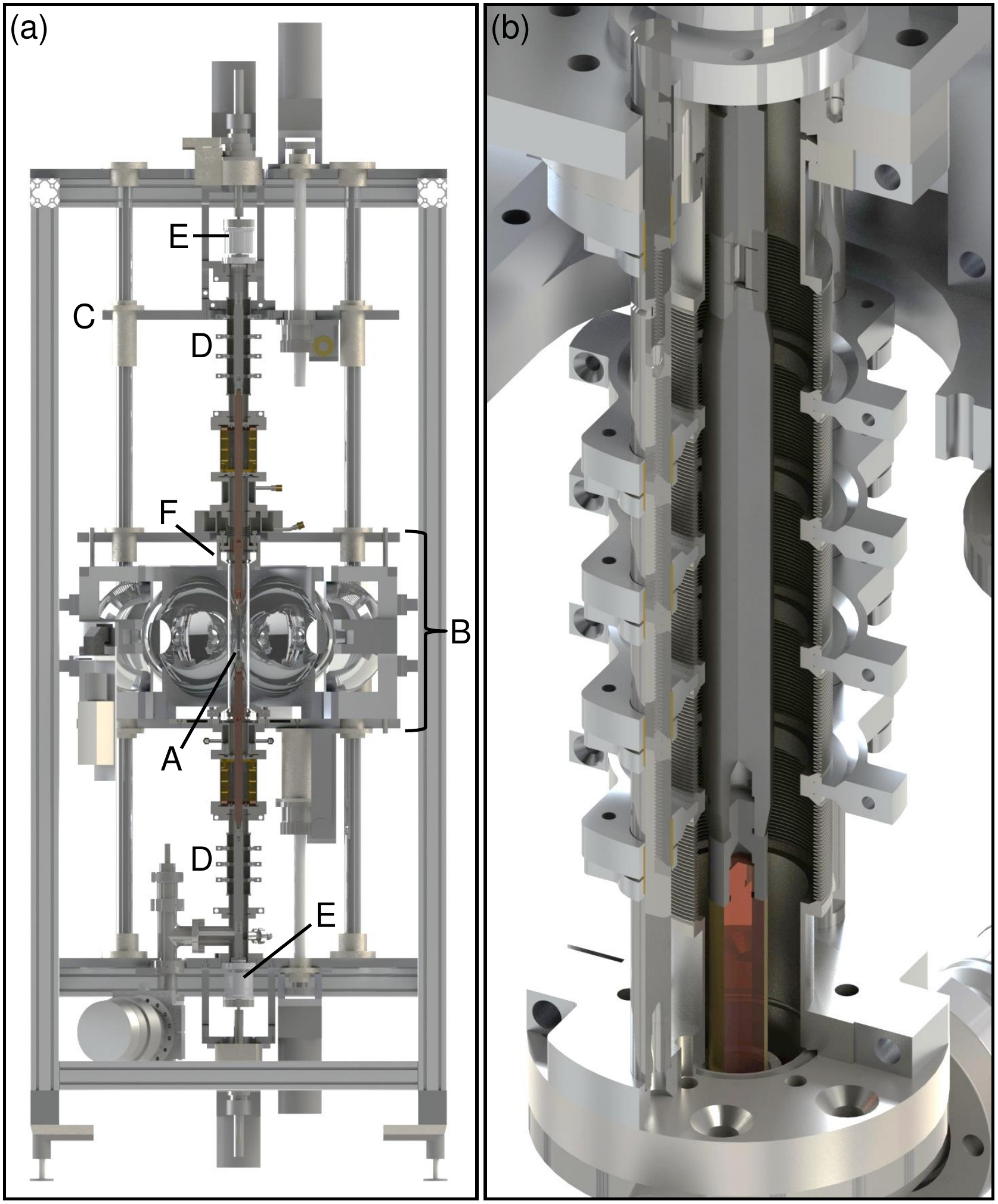}
\caption{All-metal sealed optical floating-zone furnace. (a)~Total view. (A)~Sample, (B)~mirror stage, (C)~upper stage, (D)~metal bellows, (E)~rotation feedthrough, (F)~quartz-glass assembly. (b)~Cut-away view of the high-pressure metal bellows system with guiding at ambient.}
\label{figure06}
\end{figure}

The actual single-crystal growth in our materials preparation chain is typically carried out by means of the optical floating-zone method. We use a commercial four-mirror image furnace from Crystal Systems Corporation\cite{CSC} that was, however, completely refurbished to be all-metal sealed.\cite{2011:Neubauer:RevSciInstrum} In the following, we account for operational aspects and describe further improvements of the design beyond those reported in Ref.~\onlinecite{2011:Neubauer:RevSciInstrum}.

In the image furnace, cf.\ Fig.~\ref{figure06}(a), the light of four halogen lamps with an electrical power of up to 1.5~kW is focused on the sample~(A) by ellipsoidal mirrors. In most cases, the sample consists of two polycrystalline rods positioned above one another. The lower rod and upper rod are referred to as seed and feed, respectively. Heating by the lamps is used to create a narrow molten zone at the point where both rods meet. For grain selection as well as single-crystal growth, this zone is subsequently made to traverse the rods with a speed of up to 18~mm/h. The latter is achieved by vertically moving the mirror stage~(B). Typically, the growth is performed from bottom to top with the feed providing the material that crystallizes on the seed. If available, the use of an oriented single-crystalline seed is preferred. The rate at which the sample material is added to the molten zone may be regulated by moving the upper stage~(C) in turn allowing to control the diameter of the crystal grown. Metal bellows~(D) provide the necessary room for translation. The seed and the feed may be rotated individually in either direction at up to 56~rpm. The mirror stage and those parts of the vacuum chamber close to the latter are water-cooled.

As reported in Ref.~\onlinecite{2011:Neubauer:RevSciInstrum}, magnetically coupled rotary feedthroughs~(E) and Helicoflex\cite{Helicoflex} HN 130 o-rings replacing the original viton o-rings have been used to achieve all-metal sealing and bakeability. The quartz glass inside the mirror stage has been embedded into a customized assembly~(F) utilizing Helicoflex HNV 290 P seals. Additionally, the water cooling of the furnace has been optimized to permit the efficient removal of remaining water using pressurized air prior to baking out. For a detailed description of the procedure of assembling and disassembling the furnace, we refer to Ref.~\onlinecite{2011:Neubauer:RevSciInstrum}.

Recently, the vacuum chamber of the furnace was modified further. In particular, the original metal bellows have been replaced by high-pressure metal bellows made of AM 350 stainless steel, designed by COMVAT\cite{Comvat}. As part of this change, we redesigned essentially all parts of the vacuum chamber using standard CF copper gaskets instead of the Helicoflex HN 130 o-rings. As depicted in Fig.~\ref{figure06}(b), the new bellows are stabilized from the outside thereby increasing the pumping cross section in order to improve the pumping speed. This way, we reduced the time required for reaching the final pressure by roughly a factor of two. Using a turbomolecular pump backed by an oil-free scroll pump in combination with a new set of heating jackets, we are able to reach final pressures in the range of $10^{-9}$~mbar within three days. The latter represents the precondition for the application of a high-purity argon environment at pressures of up to 5~bar.

\subsection{Annealing furnaces}

Post-growth treatment of samples in the form of controlled annealing is a well established technique to relieve strain, remove defects, or improve compositional homogeneity. Annealing of polycrystalline material may promote the formation of large grains, most notably in compounds forming congruently. Moreover, carefully chosen annealing routines, sometimes combined with a quenching of the sample, may stabilize certain compositional or structural phases. Examples include single-crystal neodymium\cite{1971:Tonnies:JCrystGrowth} or the weak itinerant ferromagnet Ni$_{3}$Al\cite{1983:Bernhoeft:PhysRevB, 1984:Bernhoeft:JPhysF}.

As the sample is kept at elevated temperatures over an extended period of time, annealing requires a very clean environment. For this purpose, intermetallic samples are commonly sealed under an argon atmosphere in quartz-glass ampules, while the heating is carried out by means of a tube furnace. The permeability of quartz glass, however, significantly increases with increasing temperature limiting the maximum temperature to about $1100~^{\circ}$C. To improve the quality of the annealing process, we have set up two all-metal sealed annealing furnaces, namely a resistively heated furnace based on a commercial Knudsen effusion cell and an inductively heated furnace that is connected to the cold boat furnace.

\begin{figure}
\includegraphics[width=1.0\linewidth]{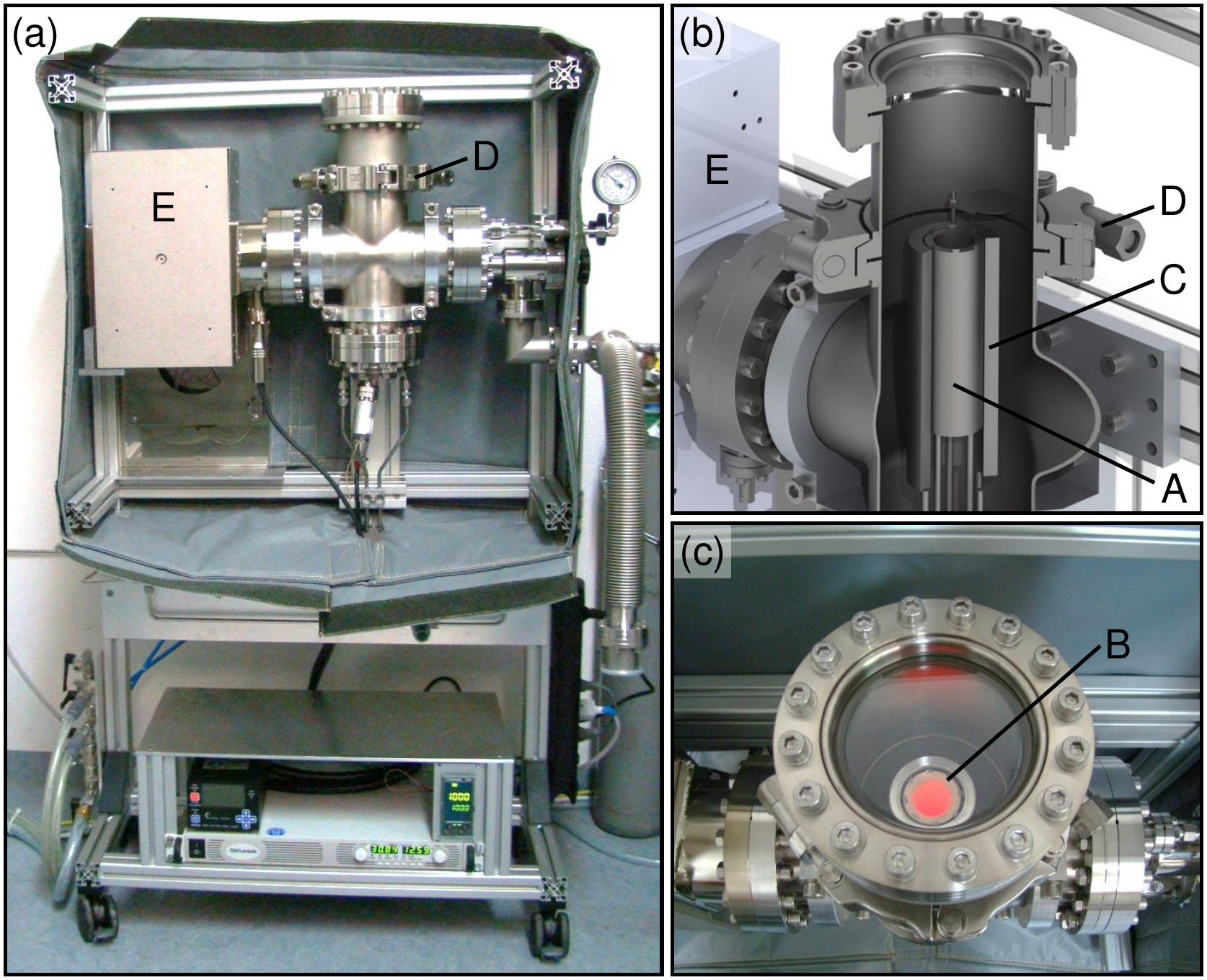}
\caption{All-metal sealed annealing furnace based on a Knudsen effusion cell. (a)~Total view. (b)~Cut-away view of the central part of the annealing furnace. (c)~View through the window while the Knudsen cell is heated. (A)~Knudsen cell, (B)~crucible out of pyrolytic boron nitride, (C)~water-cooled shroud, (D)~Quick CF flange, and (E)~ion getter pump.}
\label{figure07}
\end{figure}

Knudsen effusion cells are typically used to evaporate the educts in molecular beam epitaxy and are designed for ultra-high vacuum environments. As shown in Fig.~\ref{figure07}, in our furnace the cell~(A) is mounted on a standard CF DN40 flange. The resistive heating from a tantalum filament provides temperatures up to $1400~^{\circ}$C.\cite{MBE} The crucible~(B) is made of pyrolytic boron nitride and allows for sample volumes of ${\sim}35~\textrm{cm}^{3}$. The cell is surrounded by a bespoke water-cooled shroud~(C), also commercially available, to which we added a lid made of pyrolytic boron nitride and copper (not shown). The compact and all-metal sealed design of the vacuum chamber is based on a CF DN100 four-way cross with a Quick CF flange~(D) at the top permitting easy access for sample changes. A Gamma Vacuum TiTan L100 diode ion getter pump\cite{GammaVacuum}~(E), in combination with pre-pumping via an external turbomolecular pump backed by an oil-free scroll pump, and bake-out by means of a custom heating tent yields a vacuum of 10$^{-10}$~mbar. Up to 1.2~bar high-purity argon may be applied as an inert atmosphere.

\begin{figure}
\includegraphics[width=1.0\linewidth]{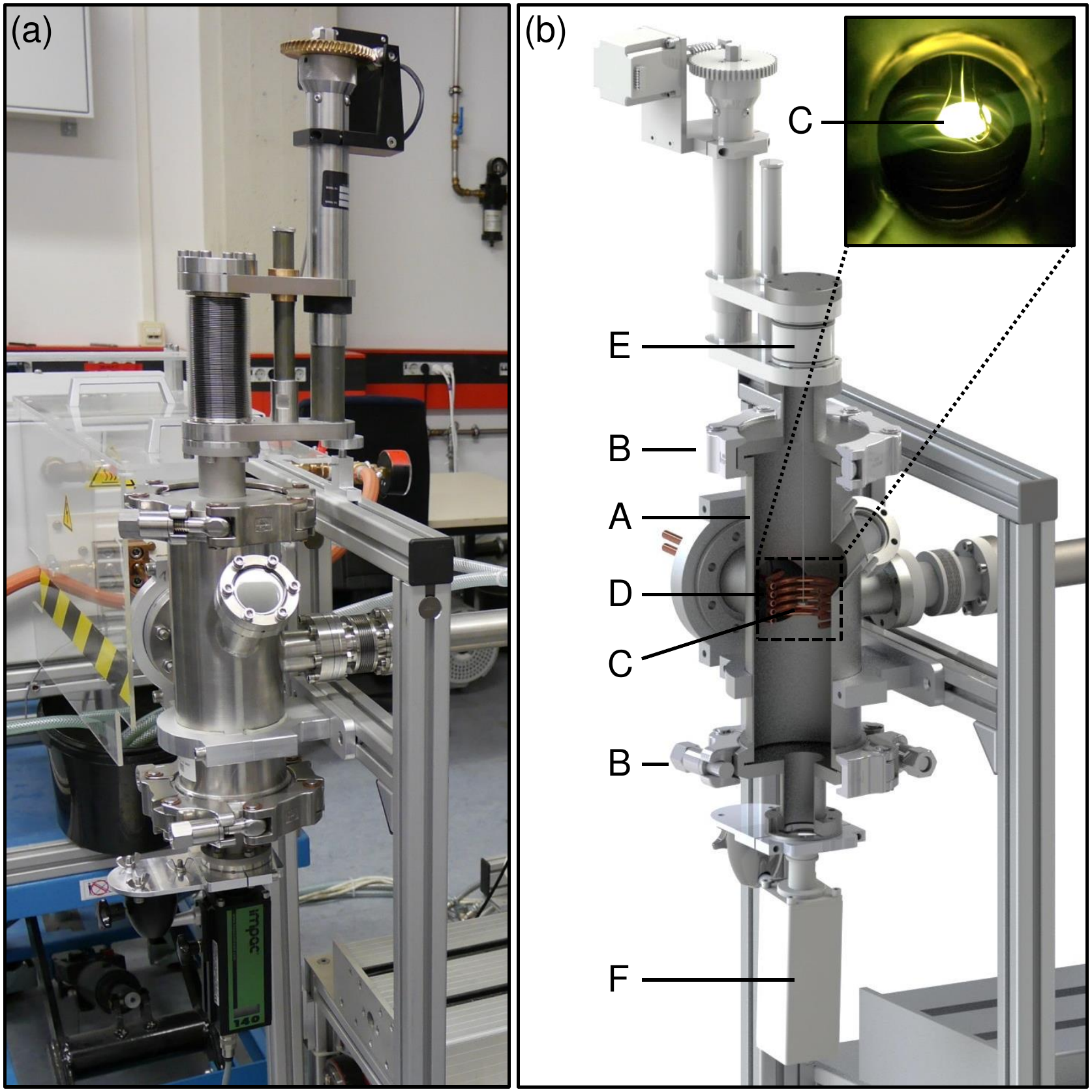}
\caption{All-metal sealed high-temperature annealing furnace. (a)~Total view. (b)~Cut-away view of the annealing furnace. (A)~Double-walled vacuum chamber, (B)~Quick CF flange, (C)~sample, (D)~RF induction coil, (E)~linear actuator, (F)~pyrometer. The inset shows a single-crystal disc of niobium suspended from a niobium wire during annealing at ${\sim}2000~^{\circ}$C.}
\label{figure08}
\end{figure}

Our inductively heated annealing furnace is depicted in Fig.~\ref{figure08} and consists of a double-walled water-cooled vacuum chamber~(A) that may replace the actual cold boat of the horizontal cold boat furnace. The latter provides pumping, gas supply, and cooling water. Two VACOM Quick CF DN100 ports~(B) tightened by clamping chains allow for easy access. The sample~(C), which is suspended from a thin wire ensuring minimal heat flux, is heated by means of RF induction using a built-in coil~(D) and the Fives Celes generator mentioned above. The wire in turn is mounted on a linear actuator~(E) which permits to move the sample with respect to the coil. The temperature of the sample is monitored through a viewing port at the bottom of the chamber using a pyrometer~(F).

Compared to the resistively heated annealing furnace, control of the sample temperature is more challenging in the inductively heated furnace. In case of sufficient coupling of the RF field to the sample, however, it allows to reach very high temperatures exceeding ${\sim}3000~^{\circ}$C, as for instance required for the treatment of selected transition metal borides.\cite{2010:Okamoto:Book} In combination with ampules and crucibles akin to those shown in Fig.~\ref{figure03}(c), vertical Bridgman growth may be carried out.

\section{Crystal growth of $\textrm{CeNi}_{2}\textrm{Ge}_{2}$}
\label{growth}

\begin{figure}
\includegraphics[width=1.0\linewidth]{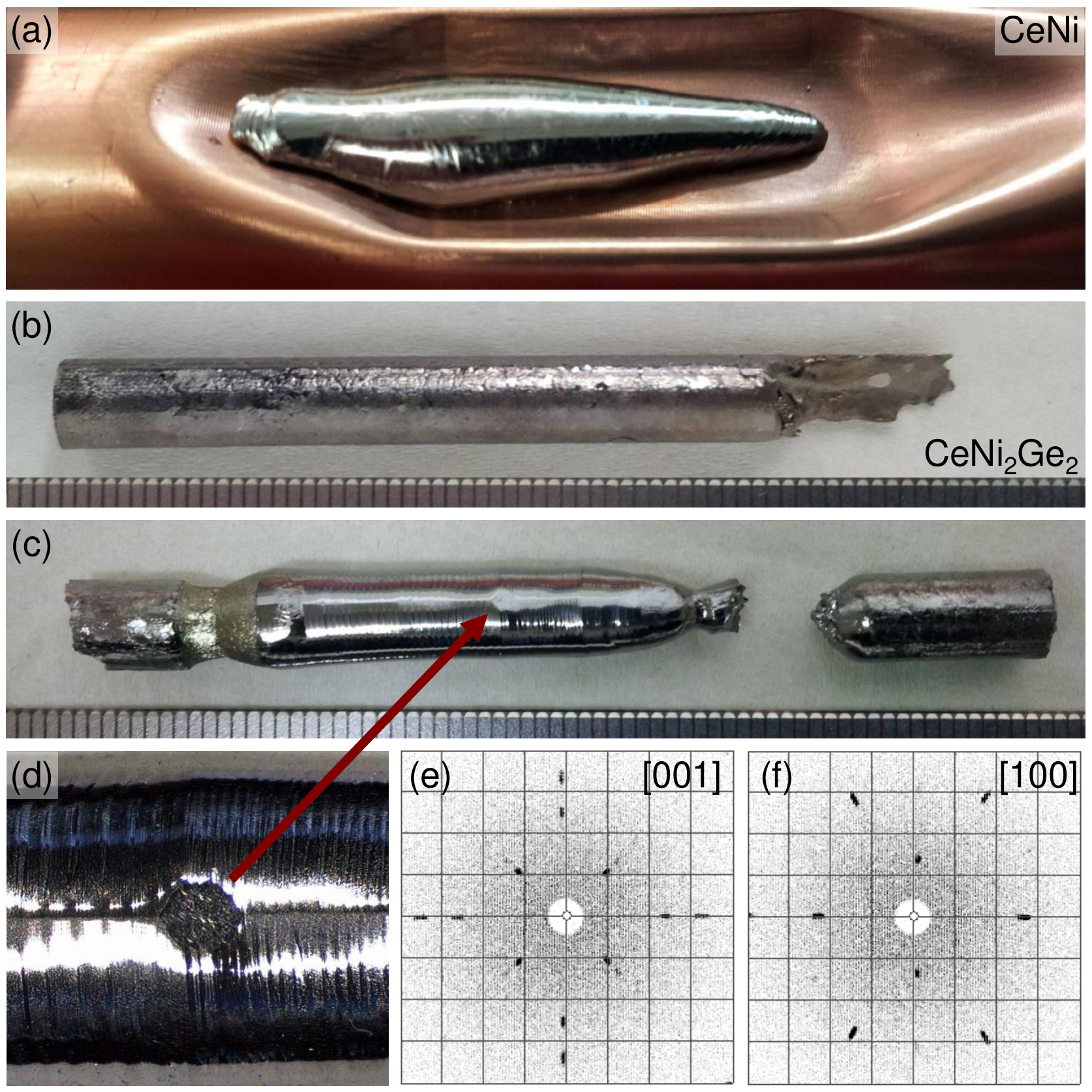}
\caption{Preparation of CeNi$_{2}$Ge$_{2}$. (a)~CeNi synthesized from high-purity elements using the cold boat furnace loaded from the argon glovebox. (b)~Polycrystalline CeNi$_{2}$Ge$_{2}$ rod cast in the rod casting furnace combining stoichiometric amounts of CeNi, Ni, and Ge. The diameter of the rod is 6~mm. (c)~Float-zoned CeNi$_{2}$Ge$_{2}$, where the growth direction was from right to left. (d)~Zoom of the central part of the single crystal. Note the shiny metallic surface and the pronounced $[001]$ facet. (e)~Fourfold Laue pattern of a $[001]$ surface. (f)~Twofold Laue pattern of a $[100]$ cut.}
\label{figure09}
\end{figure}

In order to illustrate the performance of our material preparation chain, we describe in the following the single-crystal growth of CeNi$_{2}$Ge$_{2}$. This rare-earth compound crystallizes in the tetragonal ThCr$_{2}$Si$_{2}$ structure with space group $I4/mmm$.\cite{1969:Rieger:MonatshChem} It is a heavy-fermion metal that exhibits non-Fermi liquid behavior and is discussed as being close to an antiferromagnetic quantum critical point at ambient pressure.\cite{1988:Knopp:JMagnMagnMater, 1996:Sato:JMagnMagnMater, 1996:Steglich:JPhysCondensMatter, 1997:Aoki:JPhysSocJpn, 1999:Knebel:PhysRevB, 1999:Gegenwart:PhysRevLett, 1999:Kambe:JLowTempPhys, 2000:Grosche:JPhysCondensMatter, 2000:Koerner:JLowTempPhys, 2000:Fak:JPhysCondensMatter, 2003:Kuchler:PhysRevLett, 2003:Gegenwart:JLowTempPhys} A drop of the resistivity below ${\sim}0.1$~K reported by several groups has been attributed to incipient superconductivity.\cite{1999:Gegenwart:PhysRevLett, 2000:Gegenwart:PhysicaB, 2000:Grosche:JPhysCondensMatter, 2000:Steglich:PhysicaC, 2000:Braithwaite:JPhysCondensMatter, 2006:Kawasaki:JPhysSocJpn}

A large number of studies suggests that the low-temperature physical properties of CeNi$_{2}$Ge$_{2}$, such as the electrical resistivity or the specific heat, sensitively depend on the detailed composition and the crystalline quality.\cite{1999:Gegenwart:PhysRevLett, 2000:Gegenwart:PhysicaB, 2000:Steglich:PhysicaB, 2003:Cichorek:ActaPhysPolB, 2006:Kuchler:PhysicaB, 2010:Bergmann:PhysStatusSolidiB} In these studies, samples are typically discriminated in terms of their residual electrical resistivity at low temperatures, $\rho_{0}$, or their residual resistivity ratio (RRR), i.e., the room-temperature resistivity divided by $\rho_{0}$. As the residual resistivity arises from the scattering by impurities, a low $\rho_{0}$ and a high RRR, respectively, are associated with high sample quality. Remarkably, the lowest residual resistivity value of $0.17~\mu\Omega\mathrm{cm}$ was reported for polycrystalline samples prepared from a starting composition Ce$_{1+x}$Ni$_{2+y}$Ge$_{2-y}$ with $x = 0.005$ and $y = 0.025$,\cite{2000:Gegenwart:PhysicaB} while single crystals typically show values of $\rho_{0}$ an order of magnitude larger.\cite{1983:Schneider:SolidStateCommun, 1996:Fukuhara:JPhysSocJpn, 2000:Fak:JPhysCondensMatter, 2000:Braithwaite:JPhysCondensMatter, 2003:Kuchler:PhysRevLett, 2010:Bergmann:PhysStatusSolidiB} This observation may be related to the ternary metallurgical phase diagram, where a germanium-rich impurity phase may be expected to form in polycrystals prepared from stoichiometric starting composition.\cite{1996:Salamakha:JAlloyCompd, 2010:Bergmann:PhysStatusSolidiB} The formation of this impurity phase in turn may be suppressed by a small excess of nickel. Many of the complexities of the ternary phase diagram, however, may be avoided by float-zoning which allows to start from a stoichiometric composition.

For the crystal growth of CeNi$_{2}$Ge$_{2}$ we used 5N5 high-purity cerium\cite{Ames}, 4N5 nickel, and 6N germanium. As a first step, stoichiometric amounts of cerium and nickel for the synthesis of CeNi\cite{2009:Okamoto:JPhaseEquilDiffusion} were prepared in the glovebox. CeNi, unlike elemental cerium, appears to be not sensitive to oxygen or moisture. Using the metal bellows load-lock at the glovebox, both elements were placed in the horizontal cold boat system. After pumping and baking the latter to $10^{-10}$~mbar, the system was filled with 1.1~bar high-purity argon. The specimens were molten by means of RF induction producing an ingot of CeNi as shown in Fig.~\ref{figure09}(a). Neither here nor in the following steps of the crystal growth process evaporative losses were observed.

Second, using the rod casting furnace, the CeNi ingot was combined with appropriate amounts of nickel and germanium. After pumping and baking to $10^{-8}$~mbar and flooding with 1.7~bar high-purity argon, the educts were melted by means of RF induction yielding a polycrystalline sample of stoichiometric CeNi$_{2}$Ge$_{2}$. The sample was flipped and remelted several times in order to enhance compositional homogeneity, before being cast into a polycrystalline feed rod of 6~mm diameter and ${\sim}50$~mm length, shown in Fig.~\ref{figure09}(b). The abovementioned steps were repeated a second time with a smaller amount of material to cast a seed rod of ${\sim}20$~mm length.

Third, the feed and the seed rod were mounted in the image furnace, where great care was exercised to achieve perfect alignment. The image furnace was thoroughly pumped and baked out to $10^{-9}$~mbar, before being flooded with high-purity argon at a pressure of 3~bar. Finally, the polycrystalline rods were optically float-zoned at a rate of 5~mm/h while counter-rotating feed and seed at 6~rpm. At the beginning of the growth process, necking was performed to promote grain selection. The float-zoned ingot, see Figs.~\ref{figure09}(c) and \ref{figure09}(d), displays a shiny metallic surface and pronounced facets. Already for the initial section of the ingot following the necked part, Laue X-ray diffraction indicated a single grain across the entire cross-section of the ingot. The facets were oriented perpendicular to the tetragonal $[001]$ axis while the growth direction is close to the $[100]$ axis. Typical Laue pictures are shown in Figs.~\ref{figure09}(e) and \ref{figure09}(f). Note that the yellowish taint of the surface of the final molten zone may hint at a tiny amount of impurities gathering in the molten zone during growth, most likely traces of oxygen forming a thin layer of Ce$_{2}$O$_{3}$ or CeO$_{2}$. Such behavior is frequently observed and corroborates the main advantages of the float-zoning method. The concentrations of the impurities, however, were below the detection limit of energy dispersive X-ray spectroscopy.

\begin{figure}
\includegraphics[width=1.0\linewidth]{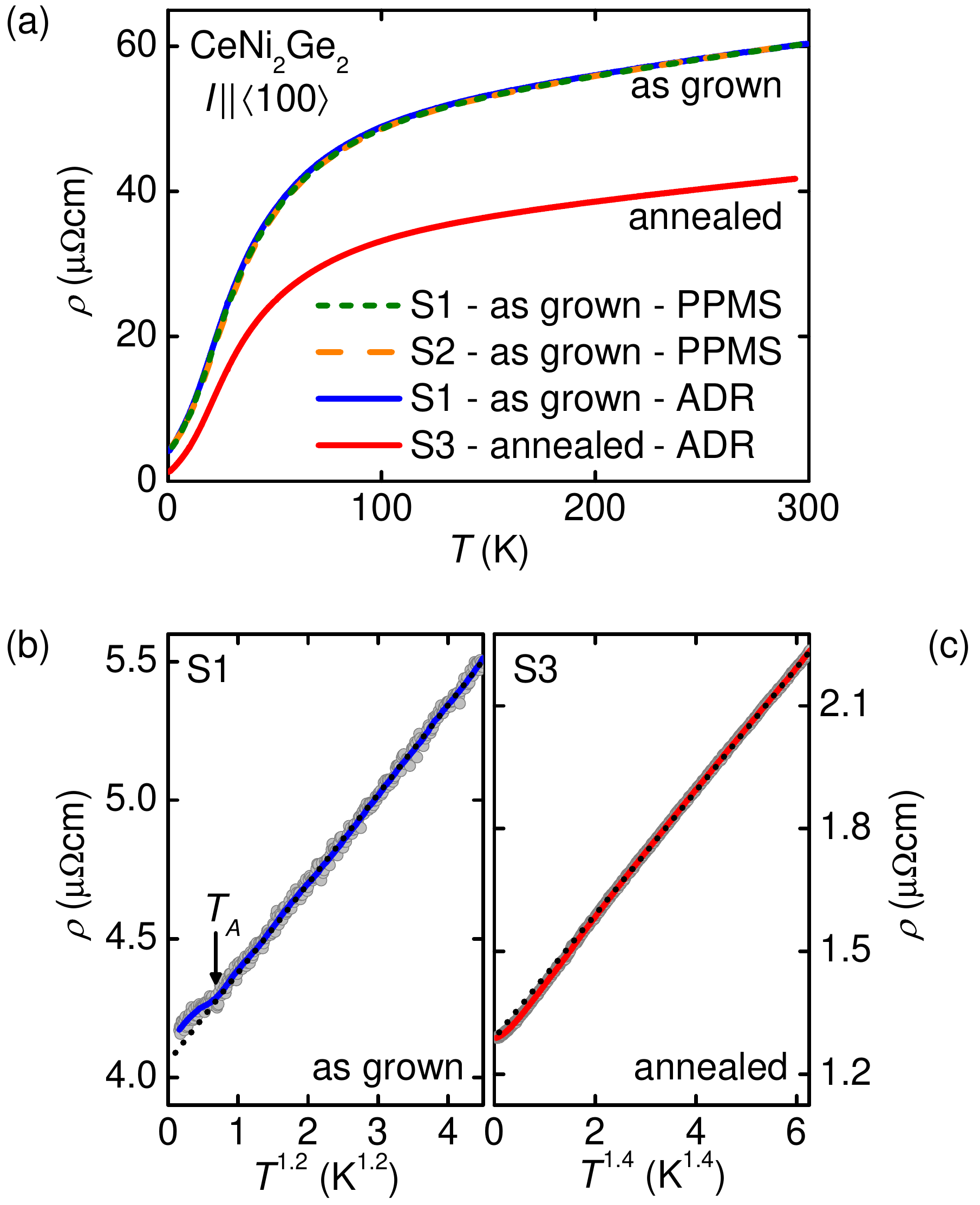}
\caption{Electrical resistivity of CeNi$_{2}$Ge$_{2}$. (a)~Electrical resistivity as a function of temperature for different samples as measured in two cryostats, PPMS and ADR. \mbox{(b),(c)}~Low-temperature resistivity of an as-grown and an annealed sample as a function of $T^{1.2}$ and $T^{1.4}$, respectively. The as-grown sample shows an anomaly at $T_{A} = 0.7$~K. The dotted lines are guides to the eye.}
\label{figure10}
\end{figure}

For measurements of the transport properties of the crystal grown, three platelets of $5 \times 1 \times 0.2~\mathrm{mm}^{3}$ were prepared with their largest surface perpendicular to $[001]$ and their long edge along $[100]$. Samples S1 and S2 were cut from the starting part and the end of the single crystal, respectively. Sample S3 was cut from the start and subsequently annealed for 100~h at $800~^{\circ}$C in the all-metal sealed furnace. Prior to this heat treatment, the furnace was pumped and baked out to $10^{-10}$~mbar before being filled with high-purity argon at a pressure of 1.1~bar.

The electrical resistivity of samples S1 and S2 were measured in a Quantum Design physical properties measurement system~(PPMS) from 300~K to 2~K. The resistivity of samples S1 and S3 were measured from room temperature down to 0.1~K in a cryogen-free system employing adiabatic demagnetization refrigeration~(ADR). In both measurement systems, a standard low-frequency lock-in technique with a four-terminal configuration was used, where the excitation currents were 5~mA and 0.1~mA, respectively.

Fig.~\ref{figure10}(a) shows the electrical resistivity of all samples as a function of temperature. With decreasing temperature the resistivity monotonically decreases, characteristic of a paramagnetic metal. In the as-grown state, data of samples S1 and S2 are in striking agreement suggesting an excellent compositional homogeneity along the float-zoned ingot. Measurements in both cryostats, PPMS and ADR, carried out on sample S1 are also in perfect agreement permitting comparison of the data. We extract a residual resistivity $\rho_{0} = 4.05~\mu\Omega\,\mathrm{cm}$ and a RRR of 14.9. After annealing the residual resistivity decreases to $\rho_{0} = 1.29~\mu\Omega\,\mathrm{cm}$ and the RRR increases to 32.6 implying a considerable improvement of sample quality. These values of $\rho_{0}$ compare with the lowest values reported for single crystals in the literature so far, where solid-state electrotransport was applied to samples grown by the Czochralski method.\cite{2000:Braithwaite:JPhysCondensMatter}

The low-temperature properties are addressed in further detail in Figs.~\ref{figure10}(b) and \ref{figure10}(c). In the as-grown state, the resistivity varies as $T^{1.2}$ with decreasing temperature until a distinct change of slope is observed at $T_{A} = 0.7$~K. A similar anomaly was also reported in Refs.~\onlinecite{2000:Steglich:PhysicaB, 2000:Steglich:PhysicaC} for several polycrystals with a slightly nickel-rich starting composition. In these studies it was pointed out that the behavior at $T_{A}$ resembles signatures at the onset of spin-density wave order in CeCu$_{2}$Si$_{2}$.\cite{1998:Gegenwart:PhysRevLett} After the annealing, the anomaly at $T_{A}$ has disappeared and the resistivity in general is best accounted for by a $T^{1.4}$ dependence. At very low temperatures, however, the measured data clearly deviate from this dependence.

\begin{figure}
\includegraphics[width=1.0\linewidth]{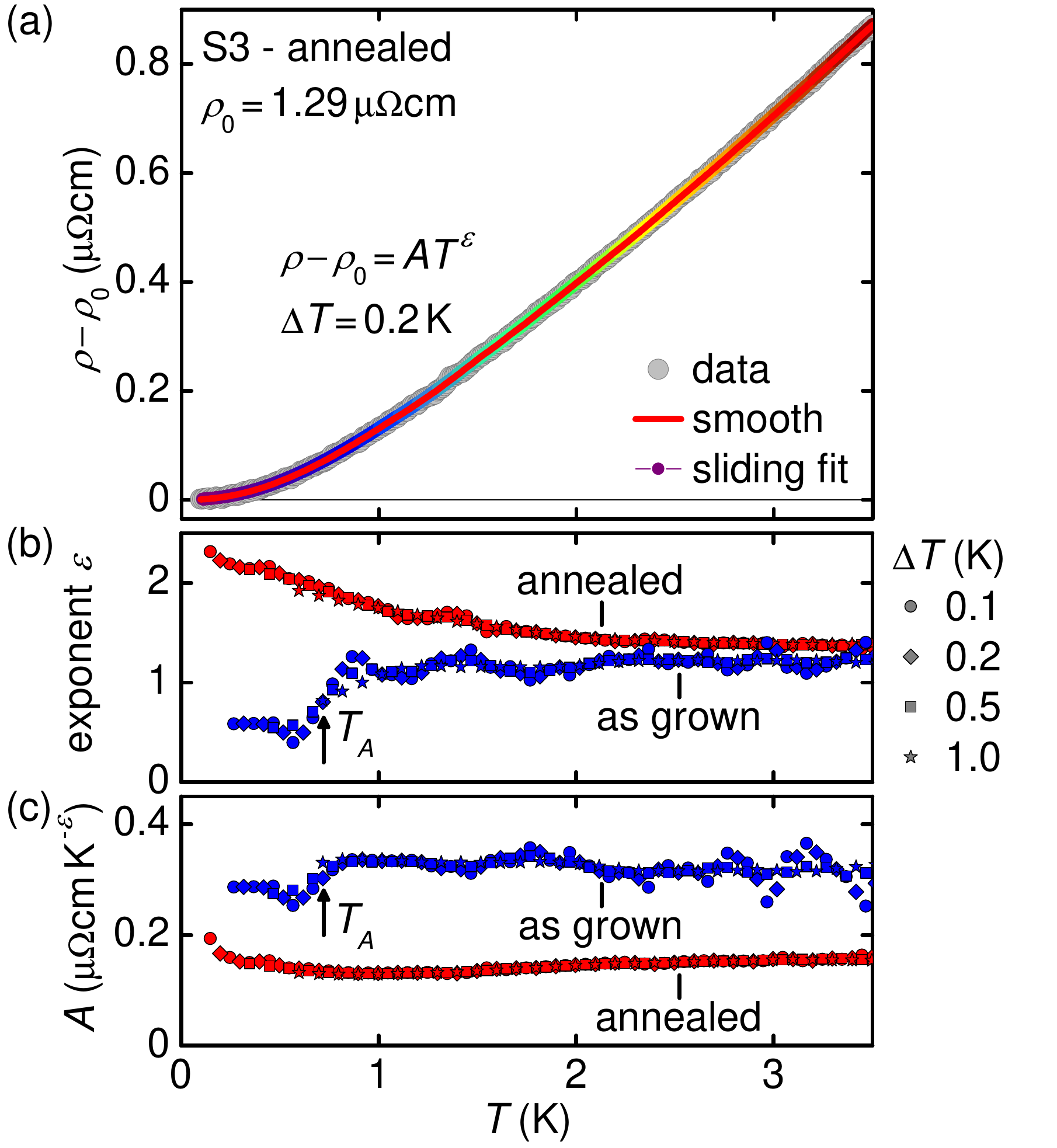}
\caption{Low-temperature resistivity of CeNi$_{2}$Ge$_{2}$. (a)~Resistivity of an annealed sample after subtraction of the residual resistivity, $\rho_{0}$. Data were fitted by a power law corresponding to $\rho - \rho_{0} = AT^{\epsilon}$. A sliding fit with a window size of $\Delta T = 0.2$~K was used. \mbox{(b),(c)}~Exponent $\epsilon$ and prefactor $A$ as a function of temperature for an as-grown and an annealed sample employing different fit window sizes.}
\label{figure11}
\end{figure}

This observation is corroborated by sliding fits as depicted in Fig.~\ref{figure11}(a). Here, after subtraction of the residual resistivity, the remaining signal is fitted with a power law dependence $\rho - \rho_{0} = A T^{\epsilon}$ in the temperature interval [$T-\Delta T/2$; $T+\Delta T/2$], where $A$ and $\epsilon$ are free parameters. Subsequently, the window is shifted by 0.1~K and the fit is repeated. The resulting exponents, $\epsilon$, and prefactors, $A$, are shown as a function of temperature in Figs.~\ref{figure11}(b) and \ref{figure11}(c). In the annealed sample, starting from $\epsilon = 1.4$ at high temperatures, the exponent monotonically increases with decreasing temperature reaching $\epsilon \approx 2.3$ at the lowest temperatures studied. The prefactor assumes values of $A \approx 0.15~\mu\Omega\,\mathrm{cm}\,\mathrm{K}^{-\epsilon}$. Window sizes between 0.1~K and 1~K yield the same results indicating the robustness of this finding. 

Our data imply that the description by means of a simple power law is insufficient for high-quality single crystals of stoichiometric CeNi$_{2}$Ge$_{2}$. Note that depending on the choice of the temperature interval, analysis with a standard power law results in exponents between 1.4 and 2.3 implicating ambiguities in the interpretation. In contrast, the as-grown sample shows stable values of $\epsilon = 1.2$ and $A \approx 0.32~\mu\Omega\,\mathrm{cm}\,\mathrm{K}^{-\epsilon}$ for $T > T_{A}$, while the fit is ill-defined for $T < T_{A}$. 

In previous reports, smaller exponents were associated with smaller residual resistivities.\cite{1999:Gegenwart:PhysRevLett, 2000:Grosche:JPhysCondensMatter, 2000:Gegenwart:PhysicaB}. This finding was corroborated by a quasi-classical treatment accounting for the interplay of strongly anisotropic scattering due to antiferromagnetic spin fluctuations and isotropic impurity scattering.\cite{1999:Rosch:PhysRevLett} Taking into account broad crossover regimes as a function of temperature in which Matthiessen's rule breaks down, the latter study suggested that the effective exponent changes from 1.5 in dirty samples to values near 1 in very clean specimens. Despite the discrepancy compared to previous reports, where a detailed discussion is beyond the scope of the technical paper presented here, our data indeed highlight that the low-temperature properties of single-crystalline CeNi$_{2}$Ge$_{2}$ depend sensitively on sample quality and are not yet fully understood.

\section{Conclusions}
\label{conclusions}

In conclusion, we presented the design of a versatile material preparation chain for intermetallic compounds either under ultra-high vacuum or under high-purity inert gas atmosphere. The chain comprises of an argon glovebox, an arc melting furnace, a horizontal cold boat furnace, a rod casting furnace, an image furnace, and two annealing furnaces. When handling materials sensitive to air, the arc melting furnace or the cold boat furnace may be connected to and loaded from the glovebox. In order to provide a growth environment as clean as possible, the furnaces are all-metal sealed, bakeable, and may be pumped to ultra-high vacuum typically representing the precondition for the application of a high-purity argon atmosphere. Besides the single-crystal growth of CeNi$_{2}$Ge$_{2}$ reported in this paper, in recent years the chain was used for the preparation of a larger number of high-quality single crystals of various intermetallic compounds. Examples include Heusler alloys,\cite{2011:Neubauer:RevSciInstrum, 2012:Neubauer:NuclInstrumMethodsPhysResA, 2015:Hugenschmidt:ApplPhysA, 2015:Schmakat:EurPhysJSpecialTopics, 2016:Bauer:RevSciInstrum} transition metal monosilicides,\cite{2010:Bauer:PhysRevB, 2010:Munzer:PhysRevB, 2010:Pfleiderer:JPhysCondensMatter, 2012:Bauer:PhysRevB, 2016:Bauer:PhysRevB, 2016:Rainer:SciRep} and various rare-earth compounds~\cite{2016:Franz:JAlloyCompd}.

\acknowledgments
We wish to thank M.\ Pfaller and the team of the Zentralwerkstatt of the Physik-Department at TUM. We gratefully acknowledge discussions with and support by P.~B\"{o}ni, R.~Bozhanova, C.~Duvinage, S.~Giemsa, S.~Gottlieb-Sch\"{o}nmeyer, M.~Horst, A.~Huxley, G.~Lonzarich, S.~Mayr, W.~M\"{u}nzer, A.~Neubauer, T.~Reimann, B.~Russ, and J.~Spallek. Financial support by the Deutsche Forschungsgemeinschaft (DFG) through grants PF393/10, PF393/11, TRR80 (From Electronic Correlations to Functionality), and FOR960 (Quantum Phase Transitions) as well as by the European Research Council (ERC) through Advanced Grant 291079 (TOPFIT) is gratefully acknowledged. G.B.\ and A.R.\ acknowledge financial support through the TUM graduate school.

\bibliography{Glovebox}

\end{document}